\documentclass[a4paper,twocolumn,
english,aps,pra,floatfix]{revtex4}
\usepackage[latin1]{inputenc}
\usepackage{amsmath}
\usepackage{babel}
\usepackage{graphics}
\usepackage{amssymb}

\begin{document}

\newcommand{\dt}{\Delta\tau}
\newcommand{\al}{\alpha}
\newcommand{\ep}{\varepsilon}
\newcommand{\ave}[1]{\langle #1\rangle}
\newcommand{\have}[1]{\langle #1\rangle_{\{s\}}}
\newcommand{\bave}[1]{\big\langle #1\big\rangle}
\newcommand{\Bave}[1]{\Big\langle #1\Big\rangle}
\newcommand{\dave}[1]{\langle\langle #1\rangle\rangle}
\newcommand{\bigdave}[1]{\big\langle\big\langle #1\big\rangle\big\rangle}
\newcommand{\Bigdave}[1]{\Big\langle\Big\langle #1\Big\rangle\Big\rangle}
\newcommand{\braket}[2]{\langle #1|#2\rangle}
\newcommand{\up}{\uparrow}
\newcommand{\dn}{\downarrow}
\newcommand{\bb}{\mathsf{B}}
\newcommand{\ctr}{{\text{\Large${\mathcal T}r$}}}
\newcommand{\sctr}{{\mathcal{T}}\!r \,}
\newcommand{\btr}{\underset{\{s\}}{\text{\Large\rm Tr}}}
\newcommand{\lvec}[1]{\mathbf{#1}}
\newcommand{\ggt}{\tilde{G}}

\newcommand{\jpsj}{J.\ Phys.\ Soc.\ Japan\ }
\newcommand{\rpp}{Rep.\ Prog.\ Phys.\ }
\newcommand{\ssc}{Solid State Commun.\ }
\def\jap{J.\ Appl.\ Phys.\ }
\def\jpcm{J.\ Phys.: Condens.\ Matt.\ }
\def\phc{Physica C\ }
\def\phe{Physica E\ }
\def\cH{{\cal H}}
\def\m{{\cal M}}

\title{A layering model for superconductivity in the Borocarbides}

\author{Thereza \surname{Paiva}, M.\ \surname{El Massalami}, and Raimundo 
R.\ \surname{dos Santos}}


\affiliation{Instituto de F\'\i sica, Universidade Federal do
Rio de Janeiro, 
21945-970 Rio de Janeiro RJ, Brazil}

\date{\today}

\begin{abstract} 
We propose a superlattice model to describe superconductivity in layered
materials, such as the borocarbide families with the chemical formul\ae\
$RT_2$B$_2$C and $RT$BC, with $R$ being (essentially) a rare earth, and
$T$ a transition metal. We assume a single band in which electrons feel a
local attractive interaction (negative Hubbard-$U$) on sites representing
the $T$B layers, while $U=0$ on sites representing the $R$C layers; the
multi-band structure is taken into account minimally through a band offset
$\varepsilon$. The one-dimensional model is studied numerically through
the calculation of the charge gap, the Drude weight, and of the pairing
correlation function. A comparison with the available information on the
nature of the electronic ground state (metallic or superconducting)  
indicates that the model provides a systematic parametrization of the
whole borocarbide family.
\end{abstract}
\maketitle

The study of Nickel borocarbides (BCs), with the general formula
$R\,$Ni$_2$B$_2$C, has given rise to a wealth of interesting information
\cite{Naga,Cava,Sieg,Muller01}. Superconductivity has been found for $R=$
Sc, Y, Ce, Dy, Ho, Er, Tm, Lu, and Th, but not for $R=$ La, Pr, Nd, Sm,
Yb, U, Gd, and Tb. In addition to the `high' critical temperature for
some compounds -- $T_c > 15$ K for $R=$ Sc, Y and Lu --, the other aspect
which has drawn attention to these materials is the coexistence, for $R=$
Dy, Ho, Er, and Tm, of superconductivity and magnetic ordering of the
$4f$ localized moments.  Structurally, the BCs are
characterized by a stacking of alternating $R$C sheets and Ni$_2$B$_2$
layers\cite{Sieg}, with both its band
structure \cite{Mattheiss94prb,Pickett94} and the near isotropic
resistivity \cite{Muller01} being consistent with three-dimensional
behaviour.

From the microscopic point of view, superconductivity in these materials
is mediated by phonons, as evidenced by specific heat \cite{sp-heat} and
isotope effect \cite{iso} measurements.  It would then appear natural to
relate the suppression of superconductivity, as $R$ and the transition
metal ($T$) are varied, in terms of the BCS parameters, $\Theta_D$,
$N(E_{\rm F})$, and $V$ (respectively, the Debye temperature, the density
of states at the Fermi level, and some measure of the electron-phonon
coupling strength): $k_BT_c=\Theta_D\exp[-1/N(E_{\rm F})V]$. Since
resistivity measurements indicate that $V$ does not vary much with $R$ in
$R\,$Ni$_2$B$_2$C \cite{lambda}, one is left primarily with $\Theta_D$ and
$N(E_{\rm F})$. While $\Theta_D$ generally increases as $R$ goes from Lu
to La, measurements of the Sommerfeld coefficient, $\gamma\propto N(E_{\rm
F})$, lead to $\gamma_{\rm Lu} \sim 2 \gamma_{\rm La}$ in the Ni series.
However, as the transition metal is varied, this simple parametrization no
longer accounts for the trend of experimental data in an unambiguous way.
Indeed, LaPt$_2$B$_2$C is a superconductor [and, for this Pt series, so
are the compounds with $R=$ Pr, Y \cite{Muller01}, and, possibly
\cite{Dhar02}, Nd], even though it has a \emph{smaller} $N(E_{\rm F})$
than non-superconducting LaNi$_2$B$_2$C: $\gamma=5$ and 8 mJ/mol K$^2$,
respectively \cite{Pt}.  Another important piece of information is
provided by studies of a closely related phase, $R\,$NiBC
\cite{Sieg,Mohammed98,RNiBC}.  The structural difference between
$R\,$Ni$_2$B$_2$C and $R\,$NiBC is the number of $R\,$C layers between the
Ni$_2$B$_2$ stacks: one in the former and two in the latter. Since these
`double-layered' materials have so far failed to display superconductivity
in both magnetic and non-magnetic systems alike, the layering structure
seems to play a crucial role in the BCs.

Considerable insight should therefore be gained by describing the
$R\,T_2$B$_2$C phenomenology in terms of zero-temperature transitions
between superconducting and normal ground states as $R$ and $T$ are
varied. With this in mind, here we propose a model based on the
expectation that superconductivity can only survive the huge 4$f$ local
moment ($\simeq 10 \mu_{\rm B}$) in, say, DyNi$_2$B$_2$C if pairing of
conduction electrons occurs solely in the NiB layers. Accordingly, we
adopt a simplified description of single-band electrons feeling an
attractive interaction only when occupying the same site on a $T_2$B$_2$
layer; electrons on sites of the $R$C layers do not interact with each
other. Our purpose here is to single out effects such as chemical
composition and layering in the suppression of superconductivity;
therefore, at this stage we will not take into account effects arising
from the $4f$ electrons of $R$, such as possible localized magnetic
moments and hybridization. The tetragonal structure suggests a further
simplification, namely, to consider a one-dimensional superlattice, an
approach which has been successfully used in interpreting some magnetic
properties of the BCs; in doing this, we still capture the essential
features of the interplay between layering along the $c$-axis, and
attractive electron-electron interactions \cite{note1}. And, finally, we
incorporate minimally the presence of actually different bands by
including a band offset between the layers; that is, we allow for the
occurrence of some charge transfer between the layers.  As we will see,
the model provides an effective parametrization, through which one is able
to systematize the data and predict which materials are likely to be
superconducting.

The above picture is realized by a one-dimensional superlattice
generalization of the attractive Hubbard model \cite{attractive,note3},
\begin{equation}
{\cal H}=\sum_{i,\,\sigma}\varepsilon_i n_{i\sigma}-
t\sum_{i,\,\sigma}\left(c_{i\sigma}^{^{\dagger}} c_{i+1\sigma}^{\phantom{\dagger}}+\text{H.c.}\right)
+ \sum_i U_i\ n_{i\uparrow}n_{i\downarrow}
\label{Ham}
\end{equation}
where the notation is standard, apart from considering position-dependent
site and correlation energy profiles: $U_i=-|U|$ and
$\varepsilon_i=\varepsilon$ for all $L_A$ sites of the attractive (i.e.,
Ni$_2$B$_2$) `layer', and $U_i=\varepsilon_i=0$ for all $L_0$ sites on the
free ($R$C) layer; the basis with $N_b=L_A+L_0$ sites is repeated
throughout the lattice. In view of the borocarbide structures, we will set
$L_A=1$, and let $L_0=1$ or 2 throughout this paper; also, from now on,
all energies will be measured in units of the hopping $t$.

Superconductivity within this model then arises due to the balance between
formation of pairs and their coherent hopping controlled by $\varepsilon$;  
e.g., as $\varepsilon\to\infty$, the system aproaches an insulating state,
since hopping will be hindered by the high energetic cost of occupying the
attractive sites. Also, in the present context we anticipate that the
strength of the attraction, $|U|$, tracks the Debye temperature,
$\Theta_D$, in the sense that for a given transition metal it increases as
the ionic radius of $R$ increases \cite{Muller01}.

\begin{table}
\caption{\label{table1} Criteria to determine the nature of the ground 
state in the attractive Hubbard superlattice, from the behavior of the 
charge gap, $\Delta$, and of the Drude weight, $D$.}
\begin{ruledtabular}
\begin{tabular}{lcr}
Nature of the state & $\Delta$ [Eq.\ (\ref{cgap})] & $D$ [Eq.\ 
(\ref{Drude})]\\
\hline
Superconducting & $\neq 0$ & $\neq 0$\\
Metallic & 0 & $\neq 0$\\
Insulating & $\neq 0$ & 0\\
\end{tabular}
\end{ruledtabular}
\end{table}

The ground state of the Hamiltonian (\ref{Ham}) has been obtained through
Lanczos diagonalization \cite{Lanczos} on lattices with $N_s$ sites and
$N_e$ electrons (thus the density $\rho=N_e/N_s$); we take closed chains
with a phase change (or magnetic flux through the ring), $\phi$, imposed
on the hopping term between sites $N_s$ and 1, in order to select
closed-shell configurations \cite{tclp00}, as well as to calculate the
Drude weight (see below). Depending on the SL configuration ($L_0=$1 or
2), and filling factor, we were able to reach lattices as large as
$N_s=24$; note, however, that not all configurations fit into all sizes
and occupations considered.

For a given SL configuration, the nature of the ground state is determined
through the calculation of the following quantities:

\noindent 
1) \emph{The charge gap,} which probes single-particle excitations, 
defined as
\begin{equation}
\Delta = E(N_s, N_e+1) + E(N_s, N_e-1) - 2 E(N_s,N_e),
\label{cgap}
\end{equation}
where $E(N_s,N^\prime)$ is the ground state energy for a chain with $N_s$
sites and $N^\prime$ electrons. 

\noindent
2) \emph{The Drude weight,} which measures the strength of the DC
conductivity peak, defined as
\begin{equation}
D=\frac{N_s}{2} \left. \frac{\partial^2  E_0}{\partial 
\phi^2}\right|_{\phi=\phi_0},
\label{Drude}
\end{equation}
where, for a given SL configuration, $E_0$ is the ground state energy for
a system with $N_e$ electrons and $N_s$ sites; $\phi_0$ is the phase value
which minimizes $E_0$.

\begin{figure}[t]
{\centering\resizebox*{3.4in}{!}{\includegraphics*{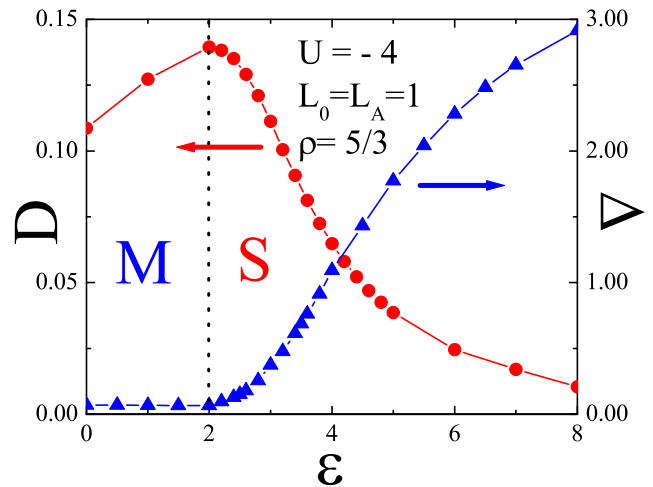}}}
\caption{\label{DDeps53} Extrapolated Drude weight (left vertical axis)  
and charge gap (right vertical axis) as functions of the site energy for
the Hubbard superlattice with $L_0=L_A=1$, $U=-4$, and an electronic
density $\rho=5/3$. The vertical dashed line separates the metallic
(M) and superconducting (S) phases.}
\end{figure}

For both $\Delta$ and $D$ we obtain a series of values for different
system sizes and extrapolate towards $N_s\to\infty$. A vanishing limit of
$\Delta$ is associated with a metallic state, whereas a non-zero value is
associated with either an insulating state or with a superconducting
state; indeed, in the case of the homogeneous (i.e., non-layered)  
attractive Hubbard model, it has been exactly shown that the
single-particle excitation spectrum has a gap for arbitrary densities
\cite{KO74}. A vanishing limit of $D$ is associated with an insulating
state, whereas a non-zero value is associated with a conducting state,
{\it i.e.,} metallic or superconductor.  Therefore, the nature of the
ground state is determined by the limiting behaviour of \emph{both}
$\Delta$ and $D$, as summarized in Table \ref{table1}.

\noindent
3) \emph{The s-wave singlet (SS) pairing correlation function,} defined as
\begin{equation}
\label{ss}
{\cal C}(i;\ell)=\frac{1}{2}\langle c_{i+\ell\downarrow} 
c_{i+\ell\uparrow}
c^\dagger_{i\uparrow} c^\dagger_{i\downarrow}+{\rm H.c.}\rangle,
\end{equation}
which is used as a consistency check for a superconducting ground state.
Typically, ${\cal C}(i;\ell)$ should display a much slower spatial decay
in the superconducting state than in the normal state.

Our results are presented by first fixing a value of $U=-4$ (in the
intermediate coupling regime), and comparing the data for different values
of $\varepsilon$ and of the electronic density; we then let $U$ vary. In
Fig.\ \ref{DDeps53} we show the extrapolated Drude weight and charge gap
for a SL with $L_0=L_A=1$, and electronic density $\rho=5/3$. According to
Table \ref{table1}, a phase transition between a metal (M) and a
superconductor (S) takes place at $\varepsilon_{c}\simeq 2$. As expected,
Fig.\ \ref{DDeps53} shows that as $\varepsilon$ is further increased, the
system approaches an insulating state. Since the insulating (I) state is
not relevant to the borocarbides (they are metallic in their normal
state), we will not discuss the S-I transition any further.

\begin{figure}[t]
{\centering\resizebox*{3.4in}{!}{\includegraphics*{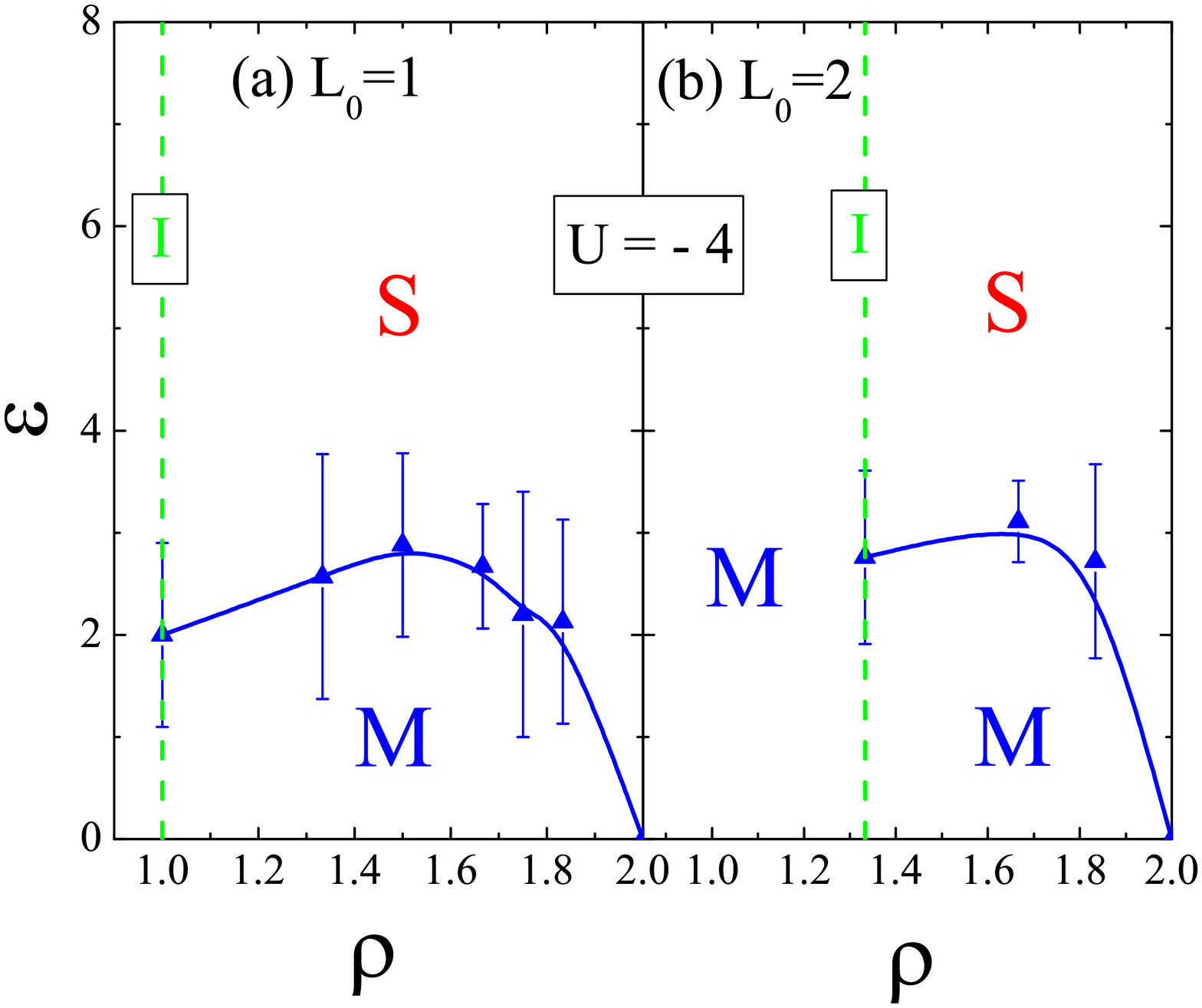}}}

\caption{
\label{eps-vs-rho} 
Phase diagram for fixed $U=-4$, and for a free layer with (a) one and (b)
two sites. Data points for $\varepsilon_c(\rho)$, and associated error
bars, come from gaussian fits to $\partial^2\Delta/\partial\varepsilon^2$.
S, M, and I stand for superconducting, metallic, and insulating phases,
respectively. Full curves guide the eye, and vertical dashed lines locate
the insulating behavior at $\rho_c$ and $\varepsilon\neq\varepsilon_c$.}
\end{figure}

Still for $U=-4$, we calculated the Drude weight and the charge gap for
other band fillings, both for $L_0=1$ ($\rho$=1, 3/2, 4/3, 7/4 and 11/6),
and for $L_0=2$ ($\rho=$4/3, 5/3, and 11/6); the analysis yields the
critical boundaries $\varepsilon_c(\rho)$ depicted in Fig.\
\ref{eps-vs-rho}. Several aspects of the phase diagrams are worth
stressing. First, a non-zero site energy is needed to stabilize the
superconducting state. Indeed, given the local (i.e., on-site) character
of the attraction, the presence of one or two free sites in between
attractive ones would normally (i.e., for $\varepsilon=0$) suppress
coherence between pairs. A repulsive site energy on the attractive sites
then has the effect of spreading the pair wave function out to neighboring
sites, increasing their overlap: pair coherence is thus recovered.
Secondly, a superconducting ground state is only stable above a critical
density, $\rho_c$, the value of which can be obtained from a strong
coupling (i.e., $\varepsilon \gg |U|\gg 1$) analysis: as one fills up an
empty lattice, electrons are first accommodated on the free sites, so
their interaction is only effective when the attractive sites start being
occupied, i.e., for densities above $\rho_{c}= 2 L_0/(L_0+L_A)$. Thirdly,
at $\rho_c$ the Drude weight (charge gap) seems to vanish (to be finite)
for all $\varepsilon$, except at $\varepsilon_c$, where it is non-zero
(zero); at $\varepsilon_c$ the system is particle-hole symmetric and
charge is evenly distributed throughout the lattice \cite{long}.

\begin{figure}[t]
{\centering\resizebox*{3.4in}{!}{\includegraphics*{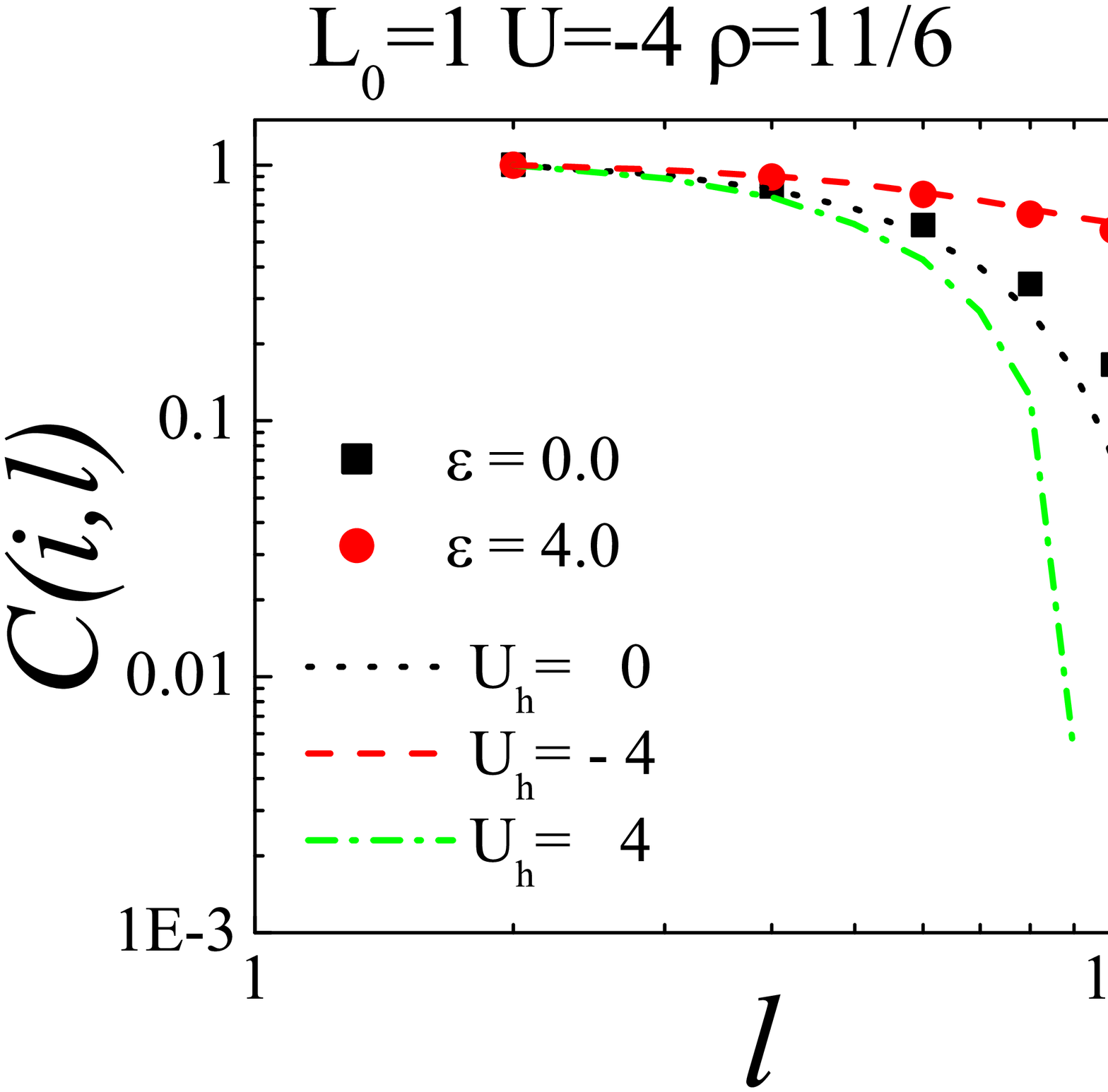}}}
\caption{\label{correlation-fn} Log-log plot of the spatial decay of the
pairing correlation function, normalized at $\ell=2$, for a 24-site
lattice and $\rho=11/6$:  dashed, dotted, and chain curves represent
data for homogeneous systems with $U_h=-4$, $0$, and 4, respectively; for
an SL with $L_0=1$ and $U=-4$, the circles and squares are data for
$\varepsilon=0$ and 4, respectively.}
\label{corr}
\end{figure}

We now illustrate the analysis of the pairing correlation function. For a
clearer comparison of the different rates of decay, in Fig.\ \ref{corr} we
eliminate oscillations due to the SL structure by plotting the
correlations at every other site; also, we normalize the functions such
that the correlation between an attractive site and its first attractive
neighbour is set to one; i.e., ${\cal C}(i,\ell=2)=1$, with $i$ chosen to
be an attractive site. Let us first consider a \emph{homogeneous} lattice
with $N_s=24$ and $\rho=11/6$. When $U_h=-4$ the system is known to be
superconducting, and, accordingly, the correlation function decays about
ten times slower than those for the corresponding non-superconducting
cases, free ($U_h=0$), and repulsive ($U_h=4$). Considering now the case
of an $L_0=1$ superlattice, also with $N_s=24$, $\rho=11/6$, and $U=-4$ we
see that for $\varepsilon=4$, the correlations decay exactly as in the
corresponding homogeneous system, whereas for $\varepsilon=0$ the rate of
decay follows more closely that of the free system. This confirms the
prediction from the analysis of $\Delta$ and $D$, which places the former
on the superconducting side of the diagram and the latter on the metallic
side (see Fig.\ \ref{eps-vs-rho}).

\begin{figure}[t]
{\centering\resizebox*{3.4in}{!}{\includegraphics*{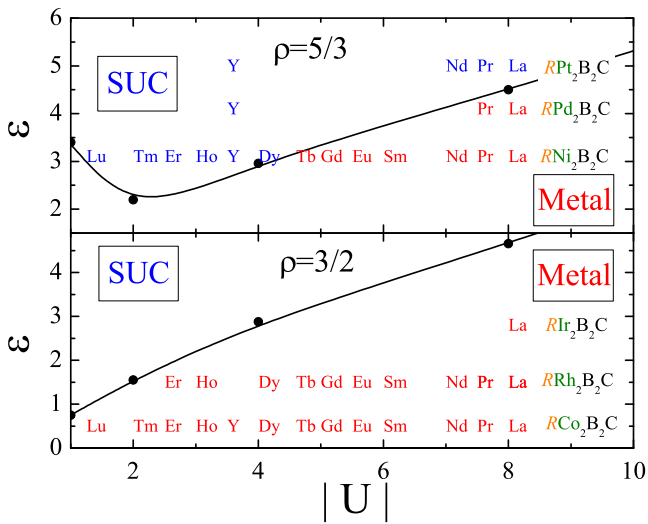}}}

\caption{
\label{eps-vs-U} 
Constant-density sections of the phase diagram
($\varepsilon$,$|U|$,$\rho$), for the superlattice with $L_0=1$ and (a)
$\rho=5/3$, and (b) $\rho=3/2$. The full curves guide the eyes through the
boundary between superconducting (SUC) and metallic phases.  The chemical
symbols correspond to $R$ in $RT_2$B$_2$C compounds; see the text.}
\end{figure}

The above procedures are repeated to determine the behavior with $U$ for
fixed densities. The resulting phase diagrams, $\varepsilon_{c}(U)$, for
$L_0=1$ are shown in Fig.\ \ref{eps-vs-U} for $\rho=5/3$ and 3/2. For
$|U|\gtrsim 2$, the site energy needed to stabilize the superconducting
phase increases with $|U|$, due to the spreading effect mentioned earlier.  
As $|U|\to 0$, a sharp rise of $\varepsilon_c$ is expected, since one
should have a metallic state for all $\varepsilon$ finite when $U=0$. For
$L_0=2$, the M-S boundary is qualitatively similar, but the S-I boundary
is more strongly affected by the presence of a second free layer,
manifested by a shift to much lower values of $\varepsilon$.

A qualitative connection with experiments on borocarbides $RT_2$B$_2$C can
now be established, based on the association of $|U|$ and $\ep$ with
measurable parameters such as the Debye temperature and the band offset.
Since $\Theta_D$ sets the scale of the BCS zero temperature gap, a
reasonable starting point is to order the $R$s according to their Debye
temperatures; these, in turn, track their ionic radii. Then, in Fig.\
\ref{eps-vs-U}(a) we first place the whole $R$Ni$_2$B$_2$C series in such
way that the S-M boundary lies between Dy and Tb; this pins $R$ to
positions on the horizontal axis \cite{note2}. Interestingly, this
parametrization subsumes the roles of $N(E_{\rm F})$, the electron-phonon
interaction and of $\Theta_D$, resolving the puzzling fact that
LaNi$_2$B$_2$C is not superconducting, in spite of the relatively large
$N(E_{\rm F})$ and $\Theta_D$.

As one changes the transition metal (keeping $R$ fixed), by going down a
$T$-column in the periodic table, the electronic density in the valence
orbitals does not change. Therefore, given our model parameters, one
expects the band offset, $\varepsilon$, to be the one most strongly
affected by this change, since it is related to the energy of the atomic
levels. Data taken from atomic energy calculations (see, for example,
\onlinecite{nist}) indicate that the energy of the higher level increases
as one goes from Ni to Pd to Pt, as well as from Co to Rh to Ir.  
Accordingly, in Fig.\ \ref{eps-vs-U}(a) the horizontal line representing
the $T=$ Pt series should be placed above the one for Ni, and in such a
way that La lies within the superconducting phase.  This resolves the
second puzzle of the borocarbides, the appearance of superconductivity as
$T$ is changed: as one moves down a column of the VIIIA group, the range
of $R$s leading to superconductivity widens.  One can therefore predict
that if a chemically stable phase of, say SmPt$_2$B$_2$C, is grown, it
should be superconducting. The Pd series suffers from a similar scarcity
of data due to chemical instabilities: apart from superconducting
YPd$_2$B$_2$C, only compounds with La and Pr have been grown so far, which
are not superconducting.  According to the phase diagram, this was already
expected, since the horizontal line representing the Pd series should lie
in between those for Ni and Pt.

We now fix $R$ and move \emph{horizontally} on the periodic table. While
the full effects of the change from, say, Ni to Co should only be captured
within a multi-band treatment of the transition metals, one expects that
the most evident feature is a decrease in band filling, possibly
accompanied by a renormalization of $\varepsilon$. Thus, we place the
borocarbides with Co, Rh and Ir in the section of the phase diagram
corresponding to a smaller $\rho$, as in Fig.\ \ref{eps-vs-U}(b).  Since
all members of the $T=$ Co series have so far failed to display
superconductivity, they should be represented by a horizontal line lying
entirely in the metallic phase; see Fig.\ \ref{eps-vs-U}(b). Many
compounds in the Rh series ($\varepsilon_\textrm{Rh} >
\varepsilon_\textrm{Co}$) have been grown, also without superconducting
members; this series is therefore represented by a line just above that
for Co. The phase diagram of Fig.\ \ref{eps-vs-U}(b) predicts that
superconductivity should be more likely for the Ir series
($\varepsilon_\textrm{Ir} > \varepsilon_\textrm{Rh}$), but this series is
even more severely plagued by chemical instabilities than the Pd and Pt
ones, and the only compound grown so far is the non-superconducting
LaIr$_2$B$_2$C.  For completeness, one should mention that the above
reasoning indicates that superconducting YRu$_2$B$_2$C should be placed in
a diagram corresponding to $\rho < 3/2$. Finally, since the double layered
materials seem to be non-superconducting, they should all lie below the
$\varepsilon_{c}$ line in the corresponding $L_0=2$ diagram.

In summary, we have proposed a superlattice model to describe
superconducting layered materials. The model is parametrized in terms of
band offset $\varepsilon$, on-site attraction $U$, electron density
$\rho$, and number of free layers $L_0$. The numerical analysis of the
one-dimensional case (at zero temperature and without local moments)
established that superconductivity is possible only above a critical
density $\rho_c(L_0)$ and for $\varepsilon > \varepsilon_c(\rho,U)$. The
model captures the essencial physics of superconductivity in the
borocarbides family, as evidenced by the available experimental
information [regarding the nature of the ground state (i.e., metallic or
superconducting)] being consistently accommodated in projections of the
phase diagram.  Consequently, one is able to predict which compounds
should become superconductors, once the barriers of chemical instability
are broken. The interplay with magnetic ordering is currently being
investigated through the addition of a Kondo-like term, which couples the
conduction electrons to the local moments. 

The authors are grateful to H Westfahl, Jr, for enlightening discussions.
Research partially supported by 
FAPERJ, CNPq, Instituto
do Mil\^enio para Nanoci\^encias (MCT), and Rede Nacional de
Nanoci\^encias (CNPq).


\begin{thebibliography}{99}
\bibitem{note3}
The attractive Hubbard model is actually more suitable to describe
short--coherence-length superconductors (which is not the case of the
BCs); nonetheless, this real-space representation of pairing makes it an
ideal framework to discuss spatial effects such as layering and disorder,
though, admittedly, at the expense of missing out on the comparison with
some specific experimental data.
\bibitem{note2} 
(a) This is a qualitative `fitting' to data, and \emph{no}
quantitative prediction for $U$ or $\varepsilon$ should be inferred.
(b) There is certainly some arbitrariness in this choice, since 
magnetic ordering should also influence the location of the boundary.
\end{thebibliography}
\end{document}